\documentclass[12pt]{article}
\usepackage{amsmath,amsfonts,amsthm,amssymb}
\usepackage{slashed}
\usepackage{xcolor}
\usepackage{graphicx}
\usepackage{epstopdf}
\usepackage{array,booktabs}
\usepackage{hyperref}
\usepackage{tabulary}
\usepackage{multirow}
\usepackage{array,arydshln}
\usepackage{bbm} 
\usepackage{cite}
\usepackage{mathtools} 

\hypersetup{linktoc = all}  
\hypersetup{pdfborderstyle={/S/U/W 0.5}}

\numberwithin{equation}{section}
\DeclareMathOperator{\Tr}{Tr}

\newcommand{\bepsilon}{\bar{\epsilon}}
\newcommand{\bsigma}{\bar{\sigma}}
\newcommand{\bpartial}{\bar{\partial}}
\newcommand{\bp}{\bar{p}}

\newcommand{\sd}{\slashed{\partial}}
\newcommand{\sD}{\slashed{D}}
\newcommand{\BRST}{\cQ_{\text{BRST}}}
\newcommand{\be}{\begin{equation}}
\newcommand{\ee}{\end{equation}}
\newcommand{\p}{\partial}

\makeatletter
\newcommand*{\letterdef@}{}
\newcommand*{\letterdef}[3]{%
  \def\letterdef@##1{\expandafter\newcommand\csname #1\endcsname{#2{##1}}}%
  \@tfor\@tempa :=#3\do{\expandafter\letterdef@\expandafter{\@tempa}}}
\makeatother
\letterdef{c#1} {\mathcal}{ABCDEFGHIJKLMNOPQRSTUVWXYZ} 
\letterdef{rm#1}{\mathrm} {dDmM} 

\begin{document}
\bibliographystyle{utphys}

\begin{titlepage}

\vskip 1.5in
\begin{center}
{\bf\Large{Localization and real Jacobi forms} }\vskip 1.9cm
{ Sujay K. Ashok${}^a$, Nima Doroud${}^{b,c}$ and Jan Troost${}^d$} \vskip
0.3in

 \emph{${}^{a}$}
\emph{ Institute of Mathematical Sciences \\
   C.I.T Campus, Taramani\\
   Chennai, India 600113\\ 
\vspace{.5cm}}

 \emph{${}^{b}$}
\emph{Perimeter Institute for Theoretical Physics \\
 Waterloo, Ontario, N2L 2Y5, Canada\\
 \vspace{.5cm}
{${}^{c}$}
Department of Physics, University of Waterloo,\\
Waterloo, Ontario N2L 3G1, Canada}

  \emph{\\${}^{d}$ Laboratoire de Physique Th\'eorique}\footnote{Unit\'e Mixte du CNRS et
     de l'Ecole Normale Sup\'erieure associ\'ee \`a l'universit\'e Pierre et
     Marie Curie 6, UMR
     8549.
}  \\
 \emph{ Ecole Normale Sup\'erieure  \\
 24 rue Lhomond \\ F--75231 Paris Cedex 05, France}
\end{center}
 \vskip 0.5in

\baselineskip 16pt

\begin{abstract}

  \vspace{.2in} 

  We calculate the elliptic genus of two dimensional abelian gauged
  linear sigma models with $(2,2)$ supersymmetry using supersymmetric
  localization. The matter sector contains charged chiral multiplets
  as well as St\"uckelberg fields coupled to the vector
  multiplets. These models include theories that flow in the infrared
  to non-linear sigma models with target spaces that are non-compact
  K\"ahler manifolds with $U(N)$ isometry and with an asymptotically linear
  dilaton direction. The elliptic genera are the modular
  completions of mock Jacobi forms that have been proposed recently
  using complementary arguments. We also compute the elliptic genera of
  models that contain multiple St\"uckelberg fields from first principles.

\end{abstract}
\end{titlepage}
\vfill\eject

\tableofcontents 

\section{Introduction}
\label{sec:intro}
There is a large class of two-dimensional conformal field theories
with ${\cal N}=(2,2)$ supersymmetry that can be described as infrared
fixed points of abelian gauge theories \cite{Witten:1993yc}.  An
interesting observable invariant under the renormalization group flow
is the elliptic genus of the theory \cite{Schellekens:1986yi,
  Witten:1986bf, Eguchi:1988vra, Kawai:1993jk,
  Witten:1993jg}. Recently, the calculation of elliptic genera of
these theories has been developed further, in particular through the
technique of localization \cite{ Gadde:2013dda, Benini:2013nda,
  Benini:2013xpa}. The results obtained so far have lead to elliptic
genera that are holomorphic Jacobi forms with weight zero and index
determined by the central charge.

In supersymmetric quantum mechanics, when there is a continuous spectrum,
the Witten index can become temperature dependent due to a difference in spectral
densities for bosons and fermions. (See e.g. \cite{Akhoury:1984pt} for a review.)
The same temperature dependence due to the continuum was observed in 
a two-dimensional index in \cite{Cecotti:1992qh}. Similarly
when the infrared fixed point exhibits a continuous spectrum, the
elliptic genus can contain a non-holomorphic contribution
\cite{Troost:2010ud} due to a difference in spectral density between
bosonic and fermionic right-moving primaries \cite{Akhoury:1984pt}\cite{Ashok:2011cy}. 
This difference is determined in terms of the asymptotic 
supercharge \cite{Akhoury:1984pt}\cite{Ashok:2013kk} and the continuum contribution is
dictated by the asymptotic geometry.
One
obtains as a result real Jacobi forms \cite{Zwegers,Zagier}
in physics \cite{Troost:2010ud,
  Eguchi:2010cb, Ashok:2011cy}. A known example of this phenomenon is
the cigar coset conformal field theory, which permits a gauged linear
sigma model description \cite{Hori:2001ax, Hori:2002cd}. The latter
includes a St\"uckelberg field linearly transforming under gauge
transformations, and rendering the two-dimensional gauge field
massive.

Thus, applying localization techniques\cite{Witten:1988ze,
  Witten:1991zz, Pestun:2007rz} to abelian two-dimensional gauge
theories including St\"uckelberg fields should lead to new features
that allow for elliptic genera that are real Jacobi forms. We will lay
bare these new features, and thereby prove a conjectured formula
\cite{Ashok:2013zka} for elliptic genera of gauged linear sigma models
containing a single St\"uckelberg field. Various consistency checks on
the conjecture were performed in \cite{Ashok:2013zka}, such as
reproducing the correct elliptic and modular properties, as well as
recuperating bound states of strings winding an isometric direction in
the target space \cite{Ashok:2013kk}. In this paper we prove and
extend these results by deriving formulas for the elliptic genera of
two-dimensional gauged linear sigma-models with multiple St\"uckelberg
fields.

This paper is organized as follows. In section \ref{sec:models}, we discuss the
gauged linear sigma models (GLSMs) that interest us. We review the infrared
geometry associated to models with a single St\"uckelberg field. We
compare and contrast with the gauged linear sigma models whose elliptic genera have already
been calculated in the literature. In section \ref{sec:genus} we show how one can
evaluate the path integral of the gauged linear sigma model by using localization in the
chiral and vector multiplet sector, and Gaussian integration in the St\"uckelberg sector. 
We perform the calculation first in a model with one St\"uckelberg field, and then generalize to models
with multiple St\"uckelberg fields. We compare with known results.
In section \ref{sec:conclusions} we conclude and point out possible applications and generalizations.

\noindent
{\it Note added:} While this paper was being prepared for publication, we received communication of \cite{Murthy:2013mya}, which contains overlapping results. 

\section{Gauged Linear Sigma Models with St\"uckelberg fields}
\label{sec:models}
In this section we review a class of gauged linear sigma models with
one St\"uckelberg field \cite{Hori:2001ax, Hori:2002cd}, and its relation to non-linear sigma models \cite{Kiritsis:1993pb}.
Next, we recall gauged linear sigma models with multiple St\"uckelberg fields.
\subsection{One St\"uckelberg field}
\subsubsection{The gauged linear sigma model}
The superspace action for the gauged linear sigma model of interest is
given by \cite{Hori:2001ax}
\begin{equation}
\label{HKLag}
S = \frac{1}{2\pi}\int d^2x d^4\theta \left[\sum_{i=1}^N\bar{\Phi_i}e^V\Phi_i + \frac{k}{4}(P+\bar{P}+V)^2 -\frac{1}{2e^2}\bar{\Sigma}\Sigma \right] \,.
\end{equation}
The chiral multiplets $\Phi_i$ have unit charge under the $U(1)$ gauge group, 
and the superfield $\Sigma$ is a twisted chiral superfield derived from the vector superfield $V$ \cite{Witten:1993yc}. 
The superfield $P$ is also a chiral multiplet with the complex scalar $p=p_1+ip_2$ as its lowest component. While
the field $p_1$ is a real uncharged non-compact bosonic field, the field $p_2$ is compact with period 
$2\pi \sqrt{\alpha'}$ and we set $\alpha'=1$ as in \cite{Hori:2001ax}. The field $P$ is charged
under the gauge group additively. It is  a St\"uckelberg field.
\subsubsection{The infrared non-linear sigma model}
With suitable linear dilaton boundary conditions \cite{Hori:2001ax},
the theory flows in the infrared to a conformal field theory which
has ${\cal N}=(2,2)$ supersymmetry and central charge
\be\label{centralcharge}
c = 3N\left(1+\frac{2N}{k}\right)\,.
\ee
To lowest order in $\alpha'$ these conformal field theories are described by a non-linear sigma model on a $2N$-dimensional K\"ahler manifold which has $U(N)$ isometry and a linear dilaton along a non-compact direction:
\begin{equation}
\begin{aligned}\label{KKL}
ds^2 &= \frac{g_N(Y)}{2}dY^2+\frac{2}{N^2g_N(Y)}(d\psi + N A_{FS})^2 + 2Yds^2_{FS}\,,
\\
\Phi &= -\frac{NY}{k} \, .
\end{aligned}
\end{equation}
The explicit form of $g_N(Y)$ was found in \cite{Kiritsis:1993pb}.

\subsection{Multiple St\"uckelberg fields}
\label{sub:multiple}
More general gauged linear sigma models exist \cite{Hori:2001ax} in
which one considers a $(U(1))^M$ gauge theory with $N$ chiral fields
$\Phi_{i}$ with charge $R_{ia}$ under the $a$th gauge group, and $M$
St\"uckelberg fields $P_a$. The superspace action is a simple
generalization of the action in \eqref{HKLag}:
\be
S = \frac{1}{2\pi}\int d^2x d^4\theta\left[ \sum_{i=1}^N \bar{\Phi}_i e^{\sum_{a}R_{ia}V_a}\Phi_i  + \sum_{a=1}^M\frac{k_a}{4}(P_a+\bar{P}_a + V_a)^2 -\sum_{a=1}^M\frac{1}{2e_a^2}|\Sigma_a|^2 \right]\,. 
\ee
The gauge transformations under the $U(1)^M$ are given by
\be
\Phi_i \rightarrow e^{i\sum_{a=1}^M R_{ia}\Lambda_a}\Phi_i \quad\text{and}\quad P_a \rightarrow P_a + i\Lambda_a \,.
\ee
The central charge of the conformal field theory to which this theory flows is given by 
\be
c=3\left(N + \sum_{a=1}^M \frac{2b_a^2}{k_a}\right) \,.
\ee
Here, $b_a$ is given by the sum over the charges of the chiral multiplets:
\be\label{ba}
b_a = \sum_{i=1}^N R_{ia} \,.
\ee

\section{Elliptic genus through localization}
\label{sec:genus}

\noindent
In this section, we compute the elliptic genera of the class of models reviewed in section \ref{sec:models}.  In the Hamiltonian formalism the elliptic genus is given by
\begin{equation}
\label{eg-h}
\chi = \Tr_{\cH_R}(-1)^F\, q^{L_0-\frac{c}{24}} \bar{q}^{\bar{L}_0-\frac{c}{24}}z^{J_0} \,,
\end{equation}
where $L_0$ and $\bar{L}_0$ are the right-moving and left-moving conformal dimensions in the CFT respectively and $J_0$ is the zero mode of the right-moving R-charge. 

We will evaluate the trace \eqref{eg-h} in the path integral formalism where the insertion of $(-1)^{F}$ amounts to imposing periodic boundary conditions for bosonic as well as fermionic fields. 
Furthermore, the insertion of $z^{J_{0}}$ twists the periodic boundary conditions in a manner that depends explicitly on the R-charge of the fields. 

Exploiting the invariance of the elliptic genus under the renormalization group flow, the computation can be carried out in the ultraviolet using the super-renormalizable gauged linear sigma model description 
\cite{Witten:1993yc, Benini:2013nda}. The R-charges of the fields in the GLSM can be read off from the explicit expression for the right-moving R-current in the GLSM realization of the 
$\cN=(2,2)$ superconformal algebra constructed in \cite{Hori:2001ax}.
Consequently, we can compute the elliptic genus by evaluating the partition function of the ultraviolet gauged linear sigma model with twisted boundary conditions using supersymmetric localization, as has been done for various 
compact models in \cite{Gadde:2013dda, Benini:2013nda, Benini:2013xpa}.

\subsection{Preliminaries}
\label{sec:preliminaries}

In what follows we carry out the path integration of the GLSM described by the action \eqref{HKLag} with twisted boundary conditions using supersymmetric localization. To avoid clutter, we present the computation
for a single chiral multiplet $\Phi$ minimally coupled, with gauge charge $q_{\Phi}=1$, to a $U(1)$ vector multiplet $V$ which is rendered massive by a single St\"uckelberg superfield $P$. 
The generalization to multiple chiral multiplets and multiple massive vector multiplets is straightforward.

After integrating over the Grassmann odd superspace coordinates, the action \eqref{HKLag} takes the form\footnote{Note that the volume form in the complex coordinates $\{w,\bar{w}\}$ takes the form $\rmd^{2}x = \frac{i}{2}\rmd^{2}w $.}
\begin{equation}
\label{action_111}
  S=\frac{i}{4\pi} \int \rmd^{2}w \left( \cL_{\text{c.m.}} + \frac{1}{2 e^{2}}\cL_{\text{v.m.}} + \frac{k}{2}\cL_{\text{St.}} \right)\,,
\end{equation}
where the chiral multiplet, vector multiplet and St\"uckelberg multiplet Lagrangians are given by
\begin{align}
  \label{cml}
  \cL_{\text{c.m.}} &= \bar{\phi}\left( -D_{\mu}^{2} + \sigma\bsigma + i\rmD \right)\phi + \bar{F}F - i\bar{\psi}\left(\sD - \sigma \gamma_{-} - \bsigma \gamma_{+} \right)\psi
    +i\bar{\psi}\lambda\phi-i\bar{\phi}\bar{\lambda}\psi \,,
  \\[5pt]
  \label{vml}
  \cL_{\text{v.m.}} &= \cF^{2} + \partial_{\mu}\sigma \partial^{\mu}\bsigma+\rmD^{2} +i \bar{\lambda} \sd \lambda \,,
  \\[5pt]
  \label{stl}
  \cL_{\text{St.}} &= \bar{G}G + \bsigma \sigma + D_{\mu}\bp D^{\mu}p + i \rmD (p+\bp) - i\bar{\chi} \sd \chi - i \bar{\lambda}\chi + i \bar{\chi} \lambda \,.
\end{align}
By $D_{\mu}$ we denote the gauge covariant derivative which acts canonically on the chiral multiplet fields while its action on the on the
St\"uckelberg scalar is given by
\begin{equation}
D_{\mu}p = \partial_{\mu}p - i A_{\mu}\,. 
\end{equation}
The action \eqref{action_111} is invariant under $\cN=(2,2)$ super-Poincar\'e transformations generated by the Dirac spinor supercharges $Q$ and $\bar{Q}$. The explicit realization of the supersymmetry algebra 
on the fields can be found in appendix \ref{app:susy}.

\subsubsection{Localization supercharge}

To compute the elliptic genus via supersymmetric localization we choose the supercharge
\begin{equation}
\label{cQ}
  \cQ = -Q_{1} - \bar{Q}_{1} \,,
\end{equation}
whose action on the fields is parametrized by the Grassmann even spinors
\begin{equation}
  \epsilon = \bar{\epsilon} = \left(\begin{array}{c} 1\\0\end{array}\right)\,.
\end{equation}
This supercharge satisfies the algebra
\begin{equation}
  \cQ^{2} = -2i \partial_{\bar{w}} + 2i \delta_{G}(A_{\bar{w}})
\end{equation}
where $\delta_{G}$ denotes a gauge transformation. One can easily show that the vector multiplet and chiral multiplet Lagrangians are, up to total derivatives, $\cQ$-exact, \emph{i.e.}
\begin{equation}
\begin{aligned}
  \cL_{\text{v.m.}} &= \cQ \cV_{\text{v.m.}} + \partial_{\mu}J_{\text{v.m.}}^{\mu}\,,
  \\
  \cL_{\text{c.m.}} &= \cQ \cV_{\text{c.m.}} + \partial_{\mu}J_{\text{c.m.}}^{\mu}\,.
\end{aligned}
\end{equation}
The explicit form of $\cV_{\text{v.m.}}$ and $\cV_{\text{c.m.}}$ can be found in appendix \ref{app:deformLag}.

In contrast to the vector and chiral multiplets, the action governing
the dynamics of the St\"uckelberg field $P$ is not globally
$\cQ$-exact\footnote{Locally, one can write the St\"uckelberg action as
  $\cQ\Lambda$, however, one can check that $\Lambda$ does not fall
  off fast enough near infinity to be in the Hilbert space of the
  theory.} \cite{Hori:2001ax}. 
This must be the case since the coefficient of the $P$-field action, $k$, appears explicitly in the expression for the central charge \eqref{centralcharge}. 
Therefore to obtain the contribution from the St\"uckelberg multiplet to the path integral via supersymmetric localization, a non-degenerate and globally $\cQ$-exact deformation term would need to be constructed.
This, however, is not necessary since the St\"uckelberg Lagrangian \eqref{stl} is quadratic, leading to a Gaussian path integral which can be explicitly carried out.

Consequently, exploiting the $\cQ$-exactness of the vector multiplet and chiral multiplet Lagrangians, we may rescale them independently by positive real numbers leaving the path integral invariant. 
While rescaling the chiral multiplet amounts to the replacement $\cL_{\text{c.m.}}\rightarrow t \cL_{\text{c.m.}}$, rescaling the vector multiplet Lagrangian is equivalent to rescaling of the Yang-Mills 
coupling $e$. In particular, we may compute the path integral in the large $t$ and $1/e^{2}$ limit, keeping the product $te^{2}$ finite. The saddle-point approximation is one-loop exact.

\subsubsection{R-charges and twisted boundary conditions}

In order to compute the path integral corresponding to the elliptic genus \eqref{eg-h}, we need to identify the charge assignments of the GLSM fields under the right moving R-symmetry. Using the explicit
expression \cite{Hori:2001ax} for the corresponding current \begin{equation}
\label{j^R}
\begin{aligned}
  j^{R}_w &=-i\left[ \bar{\psi}_{1} \psi_{1} + \frac{k}{2} \bar{\chi}_{1} \chi_{1} + \frac{i}{e^2} \bar{\sigma} \partial \sigma -i D_{w}(p-\bar{p} )\right]\,,
  \\
  j_{\bar{w}}^{R} &= -i\left[\frac{1}{2 e^2} \bar{\lambda}_{2} \lambda_{2} +  \frac{i}{e^2} \bar{\sigma} \bar{\partial} \sigma -i D_{\bar{w}} (p-\bar{p})\right]\,,
\end{aligned}
\end{equation}
yields the charge assignments 
\begin{equation}
  q^{R}_{\sigma} = q^{R}_{\lambda_{2}} = q^{R}_{\psi_{1}} = q^{R}_{\chi_{1}} =1\,,
\end{equation}
and the opposite charge for the barred fields. The zero mode of $p_2$ also carries R-charge, equal to $-\frac{1}{k}$. In addition to the dynamical fields, supersymmetry also fixes the R-charges of the auxiliary fields 
to be $q_{F}=q_{G}=1$.

The R-charges above determine the boundary conditions that need to be imposed on the GLSM path integral\footnote{This is the method followed in \cite{Ashok:2011cy} for the gauged Wess-Zumino-Witten model that describes the cigar conformal
field theory.}. 
Equivalently, the boundary conditions can be implemented via weakly gauging the right moving R-symmetry. 
This amounts to turning on a background gauge-field
\begin{equation}
a^{R} = \frac{v}{2i\tau_{2}} (\rmd w - \rmd\bar{w})
\end{equation}
for the R-symmetry with the constant parameter $v$ satisfying $z=e^{2\pi i v}$. 
Note that only the boundary condition along one cycle of the torus is affected; this will also ensure a holomorphic dependence on the variable $z$. 
The background gauge field is incorporated into the theory via gauge covariantization
\begin{equation}
  \partial_{\mu} \rightarrow \partial_{\mu} - \delta_{R}(a^{R})\,.
\end{equation}

\subsubsection{Gauge fixing and supersymmetric Faddeev-Popov ghosts}

To impose the Lorentz gauge $\partial_{\mu}A^{\mu}=0$ in the path integral in a supersymmetric way, we introduce the Grassmann odd BRST operator $\cQ_{\text{BRST}}$, the gauge fixed localization supercharge 
$\hat{\cQ} = \cQ + \cQ_{\text{BRST}}$ and the ghost and anti-ghost doublets $\{c,a_{\circ}\}$ and $\{\bar{c},b\}$ such that
\begin{equation}
\begin{aligned}
  \cQ_{\text{BRST}} &= \delta_{G}(c)\,,
  \\
  \cQ_{\text{BRST}}^{2} &= \delta_{G}(a_{\circ})\,,
  \\
  \hat{\cQ}^{2} &= -2i \bpartial + 2i \delta_{R}(a^{R}) + 2i \delta_{G}(a_{\circ})\,.
\end{aligned}
\end{equation}
This fixes the supersymmetry transformations of the ghost and anti-ghost fields up to field redefinitions\footnote{\label{fn}See appendix \ref{app:deformLag} for details.}.
Note that the vector and chiral multiplet Lagrangians are also $\hat{\cQ}$ exact by virtue of the gauge invariance of $\cV_{\text{v.m.}}$ and $\cV_{\text{c.m.}}$. We further add to the action the $\hat{\cQ}$-exact
gauge fixing term
\begin{equation}
\begin{aligned}
  \frac{1}{2e^{2}}\hat{\cQ}\cV_{\text{G.F.}} &= \frac{1}{2e^{2}} \big[ (\partial_{\mu}A^{\mu})^{2} + \left(i\partial_{\mu}A^{\mu} +b/2\right)^{2} - \bar{c}\partial_{\mu}^{2}c 
  - i \bar{c}\bpartial\left(\bar{c}+2\lambda_{1} + 2\bar{\lambda}_{1}\right)  
  \\
  &\hspace{40pt} -i b_{\circ} b - i\bar{c}_{\circ}c + i\bar{c}c_{\circ} - \bar{b}_{\circ}\left(a_{\circ}-2iA_{\bar{w}} \right) \big] \,,
\end{aligned}
\end{equation}
where we have introduced the constant ghost doublets $\{b_{\circ},c_{\circ}\}$ and $\{\bar{b}_{\circ},\bar{c}_{\circ}\}$ in order to remove the ghost zero-mode$^{\ref{fn}}$.

\subsection{Evaluation of the path integral}

The path integral that we are interested in takes the form
\begin{equation}
\label{Zsch1}
  \chi= \int\cD[\Phi,V,C,P] e^{-S_{\text{St.}} - \frac{i}{4\pi}\int\rmd^{2}w \hat{\cQ}\cV}
\end{equation}
where 
\begin{equation}
\cV = t\cV_{\text{c.m.}} + \frac{1}{2e^{2}} (\cV_{\text{v.m.}} + \cV_{\text{G.F.}}) \equiv t\cV_{\text{c.m.}} + \frac{1}{2e^{2}} \cV^{\text{G.F.}}_{\text{v.m.}}\,. 
\end{equation}
As explained in section \ref{sec:preliminaries},
the $\hat{\cQ}$-exactness of $\hat{\cQ}\cV$ ensures that the path integral is independent of the couplings $t$ and $e$. We may therefore carry out the path integration in large $t$ and $1/e^{2}$ limit, 
while keeping $te^{2}$ finite, where the saddle-point approximation is valid. 

Consequently, we first have to extract the space of saddle points of $\hat{\cQ}\cV$ which we denote by $\cM$. Explicitly, the chiral multiplet and the gauge fixed vector multiplet terms in $\hat{\cQ}\cV$ are given by
\begin{equation}
\begin{aligned}
  \hat{\cQ} \cV_{\text{c.m.}} &= \bar{F}F + D^{\mu}\bar{\phi}D_{\mu}\phi + \bar{\phi}(\bar{\sigma}\sigma +i\rmD)\phi- 2i\bar{\psi}_{2}D_{w}\psi_{2} + 2i \bar{\psi}_{1}(D_{\bar{w}} - i  a^R_{\bar{w}})\psi_{1}
    \\
    &\quad+ i \bar{\psi}_{2} \bsigma \psi_{1}  - i \bar{\psi}_{1} \sigma \psi_{2} +i \bar{\phi} ( \bar{\lambda}_{1}\psi_{2} - \bar{\lambda}_{2} \psi_{1} ) - i ( \bar{\psi}_{1}\lambda_{2} - \bar{\psi}_{2}\lambda_{1} )\phi\,,
  \\[5pt]
  \hat{\cQ}\cV^{\text{G.F}}_{\text{v.m.}} &= \partial^{\mu}A^{\nu}\partial_{\mu}A_{\nu} + \rmD^{2} + \tilde{b}^{2} + (\partial^{\mu} + i a_R^{\mu})\bsigma(\partial_{\mu} - i  a^R_{\mu})\sigma -i b_{\circ} b
  - \bar{b}_{\circ}\left(a_{\circ}-2iA_{\bar{w}}\right)
  \\
  &\quad - 2i \bar{\lambda}_{1} \bpartial \lambda_{1} + 2i \bar{\lambda}_{2} (\partial - i  a^R_{w})\lambda_{2} +\partial^{\mu} \bar{c}\partial_{\mu}c - i \bar{c}\bpartial\left(\bar{c}+2\lambda_{1} + 2\bar{\lambda}_{1}\right)
  - i\bar{c}_{\circ}c + i\bar{c}c_{\circ} \,,
\end{aligned}
\end{equation}
where $\tilde{b} = b/2 + i \partial_{\mu}A^{\mu}$. Before we look for the space of saddle points $\cM$, note that the constant ghost multiplet fields $\{c_{\circ},\bar{c}_{\circ},b_{\circ},\bar{b}_{\circ}\}$ appear as 
Lagrange multipliers and can be integrated out. This yields a delta function for the ghost zero-modes effectively removing them from the spectrum. The only remaining fermionic zero-mode is $\lambda_{1}=\lambda_{0}$, 
whereas the space of bosonic zero modes can be identified with the first De Rham cohomology of the torus and can be parametrized by a constant parameter $u$ as
\begin{equation}
\label{flatA}
  A = \frac{\bar{u}}{2i\tau_{2}}\rmd w - \frac{u}{2i\tau_{2}} \rmd\bar{w}\,.
\end{equation}

We remark that the bosonic superpartner of the fermionic zero-mode $\lambda_{0}$ is the constant mode of the vector multiplet auxiliary field, $\rmD_{0}$, and has to be treated separately.
The space of saddle-points is therefore parametrized by $\{\rmD_{0},u,\bar{u},\lambda_{0},\bar{\lambda}_{0}\}$. We normalize all bosonic and fermionic zero modes to have unit norm when Gaussian wavefunctions 
are integrated over the torus worldsheet. With this in mind, the partition function \eqref{Zsch1} reduces to the Gaussian path integral
\begin{equation}
\label{Zsch2}
  \chi = \int \frac{\rmd^{2}u}{2 i \tau_{2}} \int \rmd\rmD_{0} \int  d^{2}\lambda_{0}
  \int \cD[P] \int \hat{\cD}[eV,eC,t^{-1/2}\Phi] e^{-S_{\text{St.}}|_{\cM} - \frac{i}{4\pi}\int\rmd^{2}w \hat{\cQ}\cV|_{\text{quad }\cM} }
\end{equation}
where $\hat{\cD}[eV,eC,t^{-1/2}\Phi]$ denotes the path integral measure with the zero-modes removed. Here $\hat{\cQ}\cV|_{\text{quad }\cM}$ is the quadratic action for the fluctuations of order $e$ and order $t^{-1/2}$
for the vector multiplet and chiral multiplet fields respectively. 
The integral over $u$ is performed over the whole of the complex plane. The origin of this plane is on the one
hand the torus of holonomies of the gauge field, and on the other hand the winding modes of the compact boson
$p_2$ (the imaginary part of the St\"uckelberg field) on the toroidal worldsheet. The latter can be soaked up
into the holonomy variable $u$ such that the integral indeed covers the complex plane once.

The St\"uckelberg Lagrangian evaluated on the saddle points $\cM$ is given by
\begin{equation}
\begin{aligned}
\label{Stl}
  \cL_{\text{St.}}\Big|_{\cM} &= |G|^{2} + 4 |\partial p_{1}|^{2} + 
4 \left(\partial p_{2} - \frac{\bar{u}-v/k}{2i\tau_2}\right)  \left(\bar{\partial} p_{2} - \frac{u-v/k}{2i\tau_2}\right)\cr & \quad +2i\bar{\chi}_{1} (\bar{\partial}
    +\frac{v}{2\tau_{2}})\chi_{1} - 2i\bar{\chi}_{2} \partial \chi_{2}
    +2i\rmD p_{1} + i\bar{\chi}_{2} \lambda_{0} +i\bar{\lambda}_{0} \chi_{2}\,.
\end{aligned}
\end{equation}
Note that the kinetic term for the St\"uckelberg multiplet is not canonically normalized due to the factor of $k$ out front in equation \eqref{action_111}. 
To this end we rescale each field in the St\"uckelberg multiplet by $\sqrt{k}$. This allows us to define a canonical
measure in the path integral. With this rescaling, a few things have to be kept in mind: firstly, the periodicity of the imaginary part of the St\"uckelberg field, 
$p_2$, becomes $2\pi \sqrt{k}$. Secondly, the quadratic terms involving the zero-modes of the vector multiplet fields acquire an overall factor $\sqrt{k}$.

The first integral to carry out is over the fermionic zero modes. To perform this integral, we isolate all the terms that depend on $\lambda_{0}$:
\begin{equation}
  \int  d^2\lambda_{0} e^{\frac{1}{4\pi} \int d^2 w \left( \bar{\phi}\bar{\lambda}_{0}\psi_{2} + \bar{\psi}_{2} \lambda_{0} \phi + \sqrt{k}\bar{\chi}_{2} \lambda_{0} + 
  \sqrt{k}\bar{\lambda}_{0}\chi_{2} \right)}\,.
\end{equation}
We pause here to point out an important difference with earlier calculations of the elliptic genera of gauged linear sigma models \cite{Gadde:2013dda , Benini:2013nda , Benini:2013xpa}. This involves
the coupling of the gaugino zero modes with the fermionic partners $\chi_{2}$ of the St\"uckelberg field $p$. In the path integral over the $P$ multiplet, we also have to soak up the fermionic zero modes of
$\chi_{2}$, as can be seen from the Lagrangian in \eqref{Stl}. Therefore, on expanding the zero mode part of the Lagrangian, the only term that contributes is the quartic term in the
fermions and that leads to a factor of $k$. 

In the models with only chiral and vector multiplets \cite{Benini:2013xpa}, one obtains rather a four-point correlator involving the chiral multiplet fields. The further coupling to the $P$-multiplet
determines the fact that another correlator is to be evaluated in the chiral multiplet sector, which turns out to be just $\langle 1 \rangle$. The only coupling between the St\"uckelberg multiplet 
and the vector multiplet that remains is the coupling to the zero mode of the auxiliary field $\rmD$. Separating out this integral, the result of doing the
$\lambda_{0}$ and $\chi_{0}$ zero mode integrals we obtain
\begin{equation}
\chi =k \int \frac{\rmd^{2} u}{2 i \tau_2} \int \rmd\rmD_{0} \int \hat{\cD}[P] e^{-\int d^2 w \cL_{\text{St.}} \left\vert_{\lambda_0=\bar{\lambda}_0=0}\right.} \langle 1 \rangle_{\text{free}} \,,
\end{equation}
where the expectation value is in the chiral and vector multiplet
sector and the hat indicates that the fermionic zero mode of the
$P$-multiplet is excluded in the path integral. The free partition
function is well known and is given
by \cite{Benini:2013nda}
\be
\langle 1 \rangle_{\text{free}} = \chi_{\text{v.m.}}\, \chi_{\text{c.m.}},
\ee
where these are given by\footnote{Strictly speaking we should write Pfaffians for the fermionic path integrals. }
\begin{align}
 \chi_{\text{v.m.}} &= \frac{\hat{\det}({\bar \p})}{\det(\bar{\p}+\frac{v}{2\tau_2})}\qquad\text{and}\qquad
 \chi_{\text{c.m.}}= \frac{\det(\bar{\p}+\frac{u+v}{2\tau_2})}{\det(\bar{\p}+\frac{u}{2\tau_2})}\,.
\end{align}
See Appendix  \ref{theta} for the explicit evaluation of the chiral multiplet contribution.
The vector multiplet contribution will naturally combine with the St\"uckelberg fields.
Turning to the latter, we have a product of
functional determinants $\Delta_i$ for each of the component fields. For the field $\chi_2$, it is given by
\be
\hat{\Delta}_{\chi_2} = \hat{\det}( \p) \,.
\ee
The hat over the $\chi_2$ determinant denotes that the zero mode has
been removed. The $\chi_1$ fermion is charged under the R-current and
leads to
\be
\Delta_{\chi_1} = \det(\bar{\p}+\frac{v}{2\tau_2})\,.
\ee
Let us consider the field $p_1$, the real part of $p$. It
has a bosonic zero mode and has to be treated carefully. Taking care
of the coupling of  $p_1$ to the auxiliary field $\rmD_{0}$, we find
that
\begin{align}
\int \rmd\rmD_{0}\ \Delta_{p_1} &= \int \rmd\rmD_{0}\int \cD[p_1]e^{\int \rmd^2 w\left[- \rmD_{0}^2 + 4p_1(\p \bar{\p})p_1 -2i\sqrt{k}\rmD_{0} p_1\right]}\cr
&=\frac{1}{(\hat{\det}(\p\bar{\p}))^{\frac{1}{2}}} \int \rmd\rmD_{0} \int \rmd p_{1,0}e^{-\int d^2w (\rmD_{0}^2+2i\sqrt{k}\rmD_{0} p_{1,0})}\cr
&=\frac{1}{\sqrt{k}}\frac{1}{(\hat{\det}(\p\bar{\p}))^{\frac{1}{2}}}
\end{align}
Therefore, up to constant factors up front we obtain just the square
root of the inverse determinant. The last component field left is the
imaginary part $p_2$ of the St\"uckelberg field. This is a
periodic variable with period $2\pi\sqrt{k}$, on account of the
earlier rescaling. It is only the zero mode of this field that is
charged under the gauge field and the R-current while the non-zero
modes are uncharged. The partition function for such a
field has been reviewed in \cite{Ashok:2011cy} and is given
by
\be
\Delta_{p_2} = \frac{\sqrt{k}}{(\hat{\det}(\p\bar{\p}))^{\frac{1}{2}}} \times 
e^{-\frac{\pi k}{\tau_2}(u-\frac{v}{k})(\bar{u}-\frac{v}{k})}\,.
\ee
The factor of $\sqrt{k}$ arises from the radius of the compact
direction \cite{yellowbook}.  Note that this contribution is not
holomorphic. The non-holomorphicity arises from the momentum and
winding modes along the compact direction. The St\"uckelberg field therefore
contributes a factor
\begin{align}
\chi_{\text{St.}}&=\frac{\det(\bar{\p}+\frac{v}{2\tau_2})}{\hat{\det}(\bar{\p})}\,  
e^{-\frac{\pi k}{\tau_2}(u-\frac{v}{k})(\bar{u}-\frac{v}{k})}\,.
\end{align}
A crucial point to note is that non-zero modes of the $P$ multiplet
have combined to produce exactly the inverse of the
contribution from the vector multiplet. This is as expected from the
supersymmetric Higgs mechanism. Combining all of the
above factors, we find that the path integral 
takes the form
\be\label{chiwithc.m.}
\chi(\tau, v) = k\int \frac{d^2u}{ 2 i \tau_2}\, \chi_{\text{c.m.}}(\tau, u,v) \, e^{-\frac{k\pi}{\tau_2}
(u-\frac{v}{k})(\bar{u}-\frac{v}{k})} \,.
\ee
Using the results in Appendix \ref{theta}, one can write this as 
\be
\chi(\tau, v)  = k\int \frac{d^2u}{2 i \tau_2} \frac{\theta_{11}(\tau, u+v)}{\theta_{11}(\tau, u)} \, e^{-\frac{k\pi}{\tau_2}(u-\frac{v}{k})(\bar{u}-\frac{v}{k})} \,.
\ee
Shifting the holonomy variable $u$ by $\frac{v}{k}$ and using the rewriting the $u$-integral in terms of the variables $(s_1, s_2)$ and momentum and winding numbers\footnote{A note about
  ranges: in \cite{Ashok:2011cy}, the conventions are such that
  the gauge holonomy variables $(s_1, s_2)$ take values between $0$
  and $1$. It is possible to combine them along with the winding and
  momentum quantum numbers $(n, m)$ to obtain a complex holonomy variable $u$
  which takes values on the complex plane.} $(m,w)$, 
we obtain
\be
\chi(\tau, v) = k\int_{0}^{1} ds_1 \int_0^1 ds_2 \frac{\theta_{11}(\tau, s_1\tau+s_2+v)}{\theta_{11}(\tau, s_1\tau+s_2)} \sum_{n,m}e^{2\pi i nv} e^{-\frac{k\pi}{\tau_2}|(n+s_1)\tau+s_2 + m+\frac{v}{k}|^2} 
e^{2\pi i v_2(m+s_2+\frac{v}{k}+\tau(n+s_1))}\,.
\ee
This is the elliptic genus of the cigar conformal field theory  \cite{Ashok:2011cy},
here exhibited in the form valid for complexified chemical potentials \cite{Ashok:2014nua}.

%
%
%
%
%
%
%
%
%
%
%
%
%
%
%
%
%
%
%
%
%
%
%

\subsection{Elliptic genera for models with multiple chiral fields}

From the discussion in the preceding section, and especially equation \eqref{chiwithc.m.}, it is clear how to obtain
the elliptic genera of the models with more chiral multiplets. The
interaction Lagrangian that couples the St\"uckelberg field to the
vector multiplet remains the same; therefore the discussion regarding the
fermionic zero modes also remains the same. Consequently the
correlator to be calculated in the chiral multiplet path integral
continues to be the identity. Therefore, we include the free partition
function of a chiral multiplet in equation \eqref{chiralmultiplet} for
each of the $N$ chiral multiplets. The only difference is in the
R-charge of the St\"uckelberg field; from the discussion in
\cite{Hori:2001ax}, it is clear that the R-charge is given by
$-\frac{N}{k}$.

Putting all this together the path integral therefore is given by
\be
\chi(\tau, v) =k \int \frac{d^2u}{2i\tau_2} \left[\frac{\theta_{11}(\tau, u+v)}{\theta_{11}(\tau, u)}\right]^N \, e^{-\frac{k\pi}{\tau_2}(u-\frac{Nv}{k})(\bar{u}-\frac{Nv}{k})} \,.
\ee
This is precisely the elliptic genus that was proposed in
\cite{Ashok:2013zka}, on the basis of its modular and elliptic
properties as well as its coding of wound bound states \cite{Ashok:2013kk} in the background spacetime in \eqref{KKL}.  All properties are consistent with it being the
elliptic genus of a conformal field theory with central charge
$c=3N(1+2N/k)$. Indeed, we have now derived this fact from first principles,
through localization. As shown in \cite{Ashok:2011cy, Ashok:2013zka}, it is
also possible to define a twisted elliptic genus by including chemical
potentials for global symmetries; 
 in this case these are
the $U(1)^N$ phase rotations of each of the chiral
multiplet fields $\Phi_i$. The
resulting twisted genera take the form
\be
\chi(\tau, v, \beta_i) = k\int \frac{d^2u}{2i\tau_2} \prod_{i=1}^N\left[\frac{\theta_{11}(\tau, u+v+\beta_i)}{\theta_{11}(\tau, u+\beta_i)}\right] \, e^{-\frac{k\pi}{\tau_2}(u-\frac{Nv}{k})(\bar{u}-\frac{Nv}{k})} \,.
\ee
These twisted genera were decomposed in holomorphic and remainder contributions in \cite{Ashok:2013zka}.
We refer to \cite{Ashok:2013zka} for the calculation of the shadow 
and an interpretation of the remainder term in terms of the asymptotic geometry.

\subsection{Elliptic genera for models with multiple St\"uckelberg fields}

In subsection \ref{sub:multiple} we discussed gauged linear sigma models with gauge
groups $U(1)^M$ and $M$ St\"uckelberg fields. We specified the gauge charges
$R_{ia}$ of all the chiral fields. In order
to write the formula for the elliptic genus, we need the R-charges of
the component fields as well. These can be read off from
the R-current recorded in \cite{Hori:2001ax}. The fermions have unit R-charge while
the zero mode of the $P_a$ field has charge $-\frac{b_a}{k_a}$, where $b_a$ is given in equation \eqref{ba}.  Using
the same logic as before, one can write down the elliptic genus
of such a theory as an integral over the $M$ holonomies of the
$U(1)^M$ gauge theory:
\be
\chi(\tau, v) = \int \prod_{a=1}^M k_a\, \frac{d^2 u_a}{2 i\tau_2} \prod_{i=1}^{N} \left[\frac{\theta_{11}(\tau, \sum_{a=1}^M R_{ia}u_a + v)}{\theta_{11}(\tau, \sum_{a=1}^M R_{ia}u_a)} \right] 
e^{-\sum_{a=1}^M \frac{k_a\pi}{\tau_2}(u_a-\frac{b_a}{k_a} v)(\bar{u}_a-\frac{b_a}{k_a} v)} \,.
\ee
One can further generalize this result by including chemical
potentials for global symmetries of the model. It would also be
interesting to analyze the decomposition of this formula in terms of holomorphic contributions and non-holomorphic terms
that modularly covariantize the contributions of right-moving ground states, following \cite{Troost:2010ud,Eguchi:2010cb,Ashok:2011cy,Ashok:2013zka}

\section{Conclusions}
\label{sec:conclusions}
We have shown that in the presence of St\"uckelberg superfields, we
can still fruitfully apply the technique of localization. The dynamics
determines the observable to be calculated by localization in the
chiral and vector multiplet sectors. We have demonstrated that the appearance
of extra fermionic zero modes simplifies the observable to be
calculated. 
After applying localization to the chiral and vector
multiplet sectors, we are left with a Gaussian integration in the
St\"uckelberg sector. 
Performing this path integral, one finds that the non-zero modes of
the St\"uckelberg multiplet cancel
 the contribution from the vector
multiplet, as one would expect from the supersymmetric Higgs
mechanism. We thereby have a derivation of the elliptic genera of
gauged linear sigma models from first principles. The associated
models are non-compact and the  elliptic genera are real
Jacobi forms.

We were thus able to prove, from first principles, a formula for elliptic genera of asymptotic
linear dilaton spaces conjectured in  \cite{Ashok:2013zka}. Moreover,
we have generalized this formula to abelian gauge theories in two dimensions
with multiple St\"uckelberg fields. 

These models appear in the context of mirror symmetry in two
dimensions \cite{ Witten:1991zz, Hori:2000kt} and in the
worldsheet description of wrapped NS$5$ branes \cite{Hori:2002cd}. It
will be interesting to verify mirror symmetry at the level of the
elliptic genera. Verifications of mirror symmetry in tensor products and orbifolds
of the cigar conformal field theory and minimal models were 
performed  in \cite{Ashok:2012qy}. In order to check the
mirror duality for the genera computed in this paper, one has to calculate
elliptic genera of non-compact Landau-Ginzburg models and their
orbifolds more generally then has been done hitherto.

Applying the
calculation of these worldsheet indices to space-time string theory
BPS state counting, along the lines of \cite{Cheng:2012tq,Dabholkar:2012nd,Haghighat:2012bm,Harvey:2013mda}, would be
a further worthwhile endeavour. It would also be
interesting to find examples of non-holomorphic elliptic genera in
higher dimensions, perhaps by the addition of St\"uckelberg fields. Since the 
phenomenon of non-holomorphic contributions to indices is generic for theories
with continuous spectra, higher dimensional manifestations are likely to be found.

\section*{Acknowledgements}

We would like to thank 
Dan Israel, Suresh Nampuri and especially Jaume Gomis for insightful discussions. S.A. would also like to thank the hospitality 
of Perimeter Institute where part of this research was carried out.
N.D. would like acknowledge support from the Perimeter Institute. Research at Perimeter Institute is supported 
by the Government of Canada through Industry Canada and by the Province of Ontario through the Ministry of Research and Innovation.

\begin{appendix}

\section{Supersymmetry variations and Lagrangians}
\label{app:susy}
In this appendix we record Lagrangians and supersymmetry variations of the fields.
We follow the notations and conventions of \cite{Doroud:2012xw} regarding spinors and gamma matrices. We choose a basis such that the two-dimensional $\gamma_{\mu}$ matrices coincide with the 
Pauli matrices $\sigma^{1,2}$. The chirality matrix is given by
\be
\gamma_3 = -i\gamma^{1}\gamma^{2} = \sigma^3\,.
\ee
This allows to define projection operators
\be
\gamma_{\pm} = \frac{1}{2}(1\pm \gamma_3) \,,
\ee
which we will use in the supersymmetry variations below. With this choice, if we consider a two component Dirac spinor $\lambda$, with
\be
\lambda = \begin{pmatrix} \lambda_1 \\ \lambda_2\end{pmatrix}\,,
\ee
then the components $\lambda_1$ and $\lambda_2$ have definite chirality $\pm 1$ respectively. 

\subsection{Vector Multiplet}
\label{subsec:vm}

The vector multiplet supersymmetry transformations are given by
\begin{equation}
\begin{aligned}
  \delta \sigma &= \bepsilon\gamma_{-}\lambda - \epsilon \gamma_{+} \bar{\lambda}
  \\
  \delta \bsigma &= \bepsilon\gamma_{+}\lambda - \epsilon \gamma_{-} \bar{\lambda}
  \\
  \delta \lambda &= i \left( \sd \sigma \gamma_{+} + \sd \bsigma \gamma_{-} + \gamma^{3} \cF + i \rmD \right)\epsilon
  \\
  \delta \bar{\lambda} &= -i \left( \sd \sigma \gamma_{-} + \sd \bsigma \gamma_{+} - \gamma^{3} \cF + i \rmD \right)\bepsilon
  \\
  \delta A_{\mu} &= -\frac{i}{2}\left(\bepsilon\gamma_{\mu}\lambda + \epsilon\gamma_{\mu}\bar{\lambda}\right)
  \\
  \delta \rmD &= -\frac{i}{2} \left( \bepsilon\sd\lambda - \epsilon\sd\bar{\lambda} \right)\,.
\end{aligned}
\end{equation}
The Lagrangian governing the dynamics of the vector multiplet fields may be written as
\begin{equation}
  \cL_{\text{v.m.}} = \frac{1}{2e^{2}} \left( \cF^{2} + \partial_{\mu}\sigma \partial^{\mu}\bsigma+\rmD^{2} +i\lambda \sd \bar{\lambda}\right)
\end{equation}

\subsection{Chiral Multiplet with Minimal Coupling}

The supersymmetry transformations for a chiral multiplet with minimal coupling to the vector multiplet are
\begin{equation}
\begin{aligned}
  \delta \phi&= \bar{\epsilon}\psi
  \\
  \delta \bar{\phi}&= \epsilon\bar{\psi}
  \\
  \delta \psi&=
    i\left(\sD\phi+\sigma\phi \gamma_{+} + \bsigma \phi \gamma_{-}\right)\epsilon+\bar{\epsilon}F
  \\
  \delta \bar{\psi}&=
    i\left(\sD\bar{\phi}+\bar{\phi}\sigma \gamma_{-} + \bar{\phi}\bsigma \gamma_{+}\right)\bar{\epsilon}+\epsilon \bar{F}
  \\
  \delta F&=
    i\left(D_{\mu}\psi\gamma^{\mu}+\sigma\psi\gamma_{+} + \bsigma \psi\gamma_{-}+\lambda\phi\right)\epsilon
  \\
  \delta \bar{F}&=
    i\left(D_{\mu}\bar{\psi}\gamma^{\mu}+\bar{\psi}\sigma \gamma_{-} + \bar{\psi}\bsigma\gamma_{+}-\bar{\phi}\bar{\lambda}\right)\bar{\epsilon}\,,
\end{aligned}
\end{equation}
and the corresponding Lagrangian is given by
\begin{equation}
  \cL_{\text{c.m.}} = \bar{\phi}\left( -D_{\mu}^{2} + \sigma\bsigma + i\rmD \right)\phi + \bar{F}F - i\bar{\psi}\left(\sD - \sigma \gamma_{-} - \bsigma \gamma_{+} \right)\psi
  +i\bar{\psi}\lambda\phi-i\bar{\phi}\bar{\lambda}\psi \,.
\end{equation}

\subsection{Chiral Multiplet with St\"uckelberg Coupling}

The St\"uckelberg field is coupled to the gauge field via the covariant differentiation
\begin{equation}
\label{StueckelbergD}
  D_{\mu} p = \partial_{\mu} p - i A_{\mu} \,.
\end{equation}
The supersymmetry transformations then take the form
\begin{equation}
\begin{aligned}
  \delta p&= \bar{\epsilon}\chi
  \\
  \delta \bp &= \epsilon\bar{\chi}
  \\
  \delta \chi&=
    i\left(\sD p+\sigma \gamma_{+} + \bsigma \gamma_{-} \right)\epsilon+\bar{\epsilon} G
  \\
  \delta \bar{\chi}&=
    i\left(\sD \bp +\sigma \gamma_{-} + \bsigma \gamma_{+} \right)\bar{\epsilon}+\epsilon \bar{G}
  \\
  \delta G&=
    -i\left(\partial_{\mu}\psi\gamma^{\mu} + \lambda \right)\epsilon
  \\
  \delta \bar{G}&=
    -i\left(\partial_{\mu}\bar{\psi}\gamma^{\mu} - \bar{\lambda} \right)\bar{\epsilon}\,,
\end{aligned}
\end{equation}
and the Lagrangian is given by
\begin{equation}
  \cL_{\text{St.}} = k \left( \bar{G}G + \bsigma \sigma + D_{\mu}\bp D^{\mu}p - i\bar{\chi} \sd \chi - i \bar{\lambda}\chi + i \bar{\chi} \lambda + i \rmD (p+\bp) \right) \, .
\end{equation}

\section{Deformation Lagrangian}
\label{app:deformLag}
In this appendix, we discuss the supersymmetry variations of the fields under the localization supercharge,
the exactness of various Lagrangians, as well as the technical subtleties in the localization scheme due
to the gauge invariance of the model.

\subsection{Vector multiplets and chiral multiplets}

The supersymmetry transformation of the vector and chiral multiplet fields, including the background $R$-current, take the form
\begin{equation}
\begin{aligned}
  \cQ \sigma &= - \lambda_{2}
  \\
  \cQ \bsigma &= \bar{\lambda}_{2}
  \\
  \cQ A_{w} &= i(\lambda_{1} + \bar{\lambda}_{1})/2
  \\
  \cQ A_{\bar{w}} &= 0
  \\
  \cQ \rmD &= i \bpartial(\lambda_{1}-\bar{\lambda}_{1})
\end{aligned}
\hspace{60pt}
\begin{aligned}
  \cQ \lambda_{2} &= 2i (\bpartial - i  a^R_{\bar{w}} )\sigma
  \\
  \cQ \bar{\lambda}_{2} &= - 2i (\bpartial + i  a^R_{\bar{w}} ) \bsigma
  \\
  \cQ \lambda_{1} &= i \cF - \rmD
  \\
  \cQ \bar{\lambda}_{1} &= i \cF + \rmD
  \\
  \cQ \cF &= - \bpartial (\lambda_{1}+\bar{\lambda}_{1})
\end{aligned}
\end{equation}
and
\begin{equation}
\begin{aligned}
  \cQ \phi &= -\psi_{2}
  \\
  \cQ \bar{\phi} &= -\bar{\psi}_{2}
  \\
  \cQ \psi_{1} &= F + i \sigma \phi
  \\
  \cQ \bar{\psi}_{1} &= \bar{F} + i\bar{\phi}\bsigma
\end{aligned}
\hspace{60pt}
\begin{aligned}
  \cQ \psi_{2} &= 2i D_{\bar{w}} \phi
  \\
  \cQ \bar{\psi}_{2} &= 2i D_{\bar{w}}\bar{\phi}
  \\
  \cQ F &= -2i (D_{\bar{w}} - i a^R_{\bar{w}}) \psi_{1} + i \sigma\psi_{2} + i\lambda_{2}\phi
  \\
  \cQ \bar{F} &= -2i (D_{\bar{w}} + i a^R_{\bar{w}}) \bar{\psi}_{1} + i \bar{\psi}_{2}\bsigma - i\bar{\phi}\bar{\lambda}_{2}\,.
\end{aligned}
\end{equation}
It is straightforward to check that the Lagrangian of the vector and chiral multiplets, including the background R-current couplings, is Q-exact: if $ \tilde{\cL} = \cL_{\text{v.m.}} + \cL_{\text{c.m.}} $, 
then $\tilde{\cL}= \cQ\cV$ where
\begin{equation}
  \cV = \cV_{\text{v.m.}} + \cV_{\text{c.m.}} \,,
\end{equation}
and
\begin{align}
  \cV_{\text{v.m.}} &= \frac{1}{4 g^{2}} \left[ \bar{\lambda}_{1} (\rmD - i\cF) - \lambda_{1} (\rmD + i\cF)+ 2i \bar{\lambda}_{2} \, (\partial - i  a^R_{w} ) \sigma - 2i \lambda_{2} \, (\partial + i  a^R_{w}) \bsigma   \right]
  \\
  \cV_{\text{c.m.}} &= \frac{1}{2} \left[ \bar{\psi}_{1} (F-i\sigma \phi) + (\bar{F} - i \bar{\phi}\bsigma )\psi_{1} - 2i \bar{\psi}_{2} \, D_{w} \phi - 2i D_{w}\bar{\phi} \, \psi_{2} 
    - i \bar{\phi}(\lambda_{1} - \bar{\lambda}_{1}) \phi \right]
\end{align}

\subsection{Gauge fixing and ghosts}
\label{ghostLagrangian}

To implement the gauge fixing condition we define the (Grassmann odd) BRST operator $\BRST$ and the ghost multiplet $\{c,a\}$ such that
\begin{equation}
\begin{aligned}
  \BRST &= iq_{G} c
  \\
  \BRST^{2} &= iq_{G} a\,.
\end{aligned}  
\end{equation}
To fix the supersymmetry transformation rules for the ghost multiplet, we require that the supercharge $\hat{\cQ}=\cQ+\BRST$ satisfy the algebra
\begin{equation}
  \hat{\cQ}^{2} = -2i \bpartial - 2 q_{R} a^R_{\bar{w}} -2 q_{G} a\,.
\end{equation}
This requires the ghost field $c$ to transform as
\begin{equation}
  \hat{\cQ} c = a - 2i A_{\bar{w}}\,,
\end{equation}
while the bosonic superpartner of the ghost field, $a$, must be supersymmetric, \emph{i.e.} $\hat{\cQ}a=0$. We next define the anti-ghost multiplet $\{\bar{c},b\}$ and the constant (zero-mode)
multiplets $\{a_{\circ},c_{\circ}\}$ and $\{\bar{c}_{\circ},b_{\circ}\}$ and add to our deformation term the gauge fixing terms
\begin{equation}
\begin{aligned}
  \hat{\cQ}\cV_{\text{G.F.}} &= \frac{1}{2}\hat{\cQ} \left(\bar{c}\cG - \frac{i}{4}\bar{c}b -\bar{c}a_{\circ} + b_{\circ}c \right)
  \\
  &= \frac{1}{2}\left( \cG^{2} + (i\cG + b/2)^{2} - \bar{c}\hat{\cQ}\cG - \frac{i}{2} \bar{c}\bpartial\bar{c} + i b a_{\circ} -i \bar{c}c_{\circ} +i \bar{c}_{\circ}c + b_{\circ}(a-2iA_{\bar{w}})\right)\,,
\end{aligned}
\end{equation}
where we have used the supersymmetry transformations
\begin{equation}
\begin{aligned}
  \hat{\cQ} \bar{c} &= ib
  \\
  \hat{\cQ} b &= -2 \bpartial \bar{c}
\end{aligned}
\hspace{60pt}
\begin{aligned}
  \hat{\cQ} c_{\circ} &= 0
  \\
  \hat{\cQ} a_{\circ} &= i c_{\circ}
\end{aligned}
\hspace{60pt}
\begin{aligned}
  \hat{\cQ} \bar{c}_{\circ} &= 0
  \\
  \hat{\cQ} b_{\circ} &= i \bar{c}_{\circ}\,.
\end{aligned}
\end{equation}
In Lorentz gauge, the ghost deformation term therefore has the form
\begin{equation}
\label{Qghost}
\begin{aligned}
  \hat{\cQ}\cV_{\text{G.F.}} &= \frac{1}{2}\Big( (\p_{\mu}A^{\mu})^{2} + ( b/2+i\p_{\mu}A^{\mu})^{2} - 4\bar{c}\partial\bpartial c -i \bar{c}\bpartial\left(\bar{c} + 2\lambda_{1} + 2\bar{\lambda}_{1}\right) 
  \\ &\qquad\ + i b a_{\circ}-i \bar{c}c_{\circ} +i \bar{c}_{\circ}c + b_{\circ}(a-2iA_{\bar{w}})\Big)\,.
\end{aligned}
\end{equation}

\section{Product representation of theta functions}\label{theta}
In this appendix, we record some formulas for calculating functional determinants of free fields with
twisted boundary conditions on the torus, and their representation in terms of $\theta$ functions.
The free (twisted) path integral of the chiral multiplets which we encountered in the main text can be put in the form
\be
\chi_{\text{c.m.}}=\frac{\det(\bar{\p}+\frac{u+v}{2\tau_2})}{\det(\bar{\p}+\frac{u}{2\tau_2})}
\ee
We will diagonalize these differential operators on the torus by using the following infinite set of functions:
\be
f_{r,s}(w,\bar{w}) = \frac{1}{2i\tau_2}((r+s\tau)\bar{w}-(r+s\bar{\tau})w)\,,
\ee
where $r, s \in \mathbb{Z}$. One can check that $\Psi_{r,s}=e^{if_{r,s}}$ is single valued under the transformations
\be
w\rightarrow w+2\pi \qquad w\rightarrow w+2\pi \tau \,.
\ee
Using this basis, it is clear that the ratio of determinants takes form of an infinite product 
\begin{equation}
\chi_{\text{c.m.}}=\frac{u+v}{v}\prod_{\{r,s\}\neq\{0,0\}}\frac{\left((r+s\tau)+u+v\right)}{\left((r+s\tau)+u\right)} \, .
\end{equation}
The factor out front can be absorbed by including the $(r,s)=(0,0)$ in the infinite product. One can  check explicitly that this is a Jacobi form with a given weight and index. Using this knowledge, one can rewrite the expression as
\begin{equation}\label{chiralmultiplet}
\chi_{\text{c.m.}}=\prod_{\{r,s\}}\frac{\left((r+s\tau)+u+v\right)}{\left((r+s\tau)+u\right)} = \frac{\theta_{11}(\tau, u+v)}{\theta_{11}(\tau, u)} \, .
\end{equation}
Similar formulae are also used in \cite{Benini:2013nda}.

\end{appendix}



\begin{thebibliography}{99}
 
\bibitem{Witten:1993yc} 
  E.~Witten,
  ``Phases of N=2 theories in two-dimensions,''
  Nucl.\ Phys.\ B {\bf 403}, 159 (1993)
  [hep-th/9301042].

 
 
 \bibitem{Schellekens:1986yi}
   A.~N.~Schellekens and N.~P.~Warner,
   ``Anomalies And Modular Invariance In String Theory,''
   Phys.\ Lett.\ B {\bf 177} (1986) 317.
 
 \bibitem{Witten:1986bf}
   E.~Witten,
   ``Elliptic Genera And Quantum Field Theory,''
   Commun.\ Math.\ Phys.\  {\bf 109} (1987) 525.
 
 \bibitem{Eguchi:1988vra}
   T.~Eguchi, H.~Ooguri, A.~Taormina and S.~-K.~Yang,
   ``Superconformal Algebras and String Compactification on Manifolds with SU(N) Holonomy,''
   Nucl.\ Phys.\ B {\bf 315} (1989) 193.
 
 \bibitem{Kawai:1993jk}
   T.~Kawai, Y.~Yamada and S.~-K.~Yang,
   ``Elliptic genera and N=2 superconformal field theory,''
   Nucl.\ Phys.\ B {\bf 414} (1994) 191
   [hep-th/9306096].


 \bibitem{Witten:1993jg}
   E.~Witten,
   ``On the Landau-Ginzburg description of N=2 minimal models,''
   Int.\ J.\ Mod.\ Phys.\ A {\bf 9} (1994) 4783
   [hep-th/9304026].

 
 \bibitem{Gadde:2013dda}
   A.~Gadde and S.~Gukov,
   ``2d Index and Surface operators,''
   arXiv:1305.0266 [hep-th].
 
 
\bibitem{Benini:2013nda} 
  F.~Benini, R.~Eager, K.~Hori and Y.~Tachikawa,
  ``Elliptic genera of two-dimensional N=2 gauge theories with rank-one gauge groups,''
  arXiv:1305.0533 [hep-th].
  
 \bibitem{Benini:2013xpa}
   F.~Benini, R.~Eager, K.~Hori and Y.~Tachikawa,
  ``Elliptic genera of 2d N=2 gauge theories,''
   arXiv:1308.4896 [hep-th].


\bibitem{Akhoury:1984pt}
  R.~Akhoury and A.~Comtet,
  ``Anomalous Behavior of the Witten Index: Exactly Soluble Models,''
  Nucl.\ Phys.\ B {\bf 246} (1984) 253.

\bibitem{Cecotti:1992qh}
  S.~Cecotti, P.~Fendley, K.~A.~Intriligator and C.~Vafa,
  ``A New supersymmetric index,''
  Nucl.\ Phys.\ B {\bf 386} (1992) 405
  [hep-th/9204102].

 
 \bibitem{Troost:2010ud}
   J.~Troost,
   ``The non-compact elliptic genus: mock or modular,''
   JHEP {\bf 1006}, 104 (2010)
   [arXiv:1004.3649 [hep-th]].
 
 \bibitem{Ashok:2011cy}
   S.~K.~Ashok, J.~Troost,
   ``A Twisted Non-compact Elliptic Genus,''
   JHEP {\bf 1103 } (2011)  067.
   [arXiv:1101.1059 [hep-th]].
 

 \bibitem{Ashok:2013kk} 
   S.~K.~Ashok, S.~Nampuri and J.~Troost,
   ``Counting Strings, Wound and Bound,''
   JHEP {\bf 1304}, 096 (2013)
   [arXiv:1302.1045 [hep-th]].
  
 
  \bibitem{Zwegers}
  S. Zwegers, PhD thesis, ``Mock Theta functions'', Utrecht University, 2002.
 
  \bibitem{Zagier}
  D. Zagier, ``Ramanujan's mock theta functions and their applications
  d'apr\`es Zwegers and Bringmann-Ono'', S\'eminaire Bourbaki, 986 (2007).
 
  \bibitem{Eguchi:2010cb} 
   T.~Eguchi and Y.~Sugawara,
   ``Non-holomorphic Modular Forms and SL(2,R)/U(1) Superconformal Field Theory,''
   JHEP {\bf 1103}, 107 (2011)
   [arXiv:1012.5721 [hep-th]].

  \bibitem{Hori:2001ax}
    K.~Hori and A.~Kapustin,
    ``Duality of the fermionic 2d black hole and N = 2 Liouville theory as mirror symmetry,''
    JHEP {\bf 0108}, 045 (2001)
    [arXiv:hep-th/0104202].

 \bibitem{Hori:2002cd}
   K.~Hori and A.~Kapustin,
   ``World sheet descriptions of wrapped NS five-branes,''
   JHEP {\bf 0211} (2002) 038
   [hep-th/0203147].

 
\bibitem{Witten:1988ze} 
  E.~Witten,
  ``Topological Quantum Field Theory,''
  Commun.\ Math.\ Phys.\  {\bf 117}, 353 (1988).

\bibitem{Witten:1991zz} 
  E.~Witten,
  ``Mirror manifolds and topological field theory,''
  In *Yau, S.T. (ed.): Mirror symmetry I* 121-160
  [hep-th/9112056].
  
\bibitem{Pestun:2007rz} 
  V.~Pestun,
  ``Localization of gauge theory on a four-sphere and supersymmetric Wilson loops,''
  Commun.\ Math.\ Phys.\  {\bf 313}, 71 (2012)
  [arXiv:0712.2824 [hep-th]].

\bibitem{Ashok:2013zka} 
  S.~K.~Ashok and J.~Troost,
  ``Elliptic genera and real Jacobi forms,''
  arXiv:1310.2124 [hep-th].


\bibitem{Murthy:2013mya} 
  S.~Murthy,
  ``A holomorphic anomaly in the elliptic genus,''
  arXiv:1311.0918 [hep-th].

\bibitem{Ashok:2014nua} 
  S.~K.~Ashok, E.~Dell'Aquila and J.~Troost,
  ``Higher Poles and Crossing Phenomena from Twisted Genera,''
  arXiv:1404.7396 [hep-th].




 \bibitem{Kiritsis:1993pb}
   E.~Kiritsis, C.~Kounnas and D.~Lust,
   ``A Large class of new gravitational and axionic backgrounds for four-dimensional superstrings,''
   Int.\ J.\ Mod.\ Phys.\ A {\bf 9} (1994) 1361
   [hep-th/9308124].


  
 
  
  
\bibitem{yellowbook} 
  P.~Di Francesco, P.~Mathieu and D.~Senechal,
  ``Conformal field theory,''
  New York, USA: Springer (1997) 890 p
 
 
  
  
  
  
 

 
 
\bibitem{Hori:2000kt} 
  K.~Hori and C.~Vafa,
  ``Mirror symmetry,''
  hep-th/0002222.
 
 
 \bibitem{Ashok:2012qy} 
   S.~K.~Ashok and J.~Troost,
   ``Elliptic Genera of Non-compact Gepner Models and Mirror Symmetry,''
   JHEP {\bf 1207}, 005 (2012)
   [arXiv:1204.3802 [hep-th]].
   
\bibitem{Cheng:2012tq}
  M.~C.~N.~Cheng, J.~F.~R.~Duncan and J.~A.~Harvey,
  ``Umbral Moonshine,''
  arXiv:1204.2779 [math.RT].


\bibitem{Dabholkar:2012nd}
  A.~Dabholkar, S.~Murthy and D.~Zagier,
  ``Quantum Black Holes, Wall Crossing, and Mock Modular Forms,''
  arXiv:1208.4074 [hep-th].


\bibitem{Haghighat:2012bm}
  B.~Haghighat, J.~Manschot and S.~Vandoren,
  ``A 5d/2d/4d correspondence,''
  JHEP {\bf 1303} (2013) 157
  [arXiv:1211.0513 [hep-th]].

\bibitem{Harvey:2013mda} 
  J.~A.~Harvey and S.~Murthy,
  ``Moonshine in Fivebrane Spacetimes,''
  arXiv:1307.7717 [hep-th].
 
\bibitem{Doroud:2012xw} 
  N.~Doroud, J.~Gomis, B.~Le Floch and S.~Lee,
  ``Exact Results in D=2 Supersymmetric Gauge Theories,''
  JHEP {\bf 1305}, 093 (2013)
  [arXiv:1206.2606 [hep-th]].


 
%
 \end{thebibliography}
\end{document}